\documentclass[aps,prl,preprint,superscriptaddress]{revtex4-1}

\usepackage{graphicx}
\usepackage{amsmath,amsfonts,amssymb}
\usepackage[colorlinks=true,linkcolor=blue,citecolor=blue]{hyperref}

\begin{document}

\title{Snake Trajectories in Ultraclean Graphene p-n Junctions}


\author{Peter Rickhaus}
\author{P\'eter Makk}
\email{peter.makk@unibas.ch}
\affiliation{Department of Physics, University of Basel, Klingelbergstrasse 82, CH-4056 Basel, Switzerland}

\author{Ming-Hao Liu}
\affiliation{Institut f\"ur Theoretische Physik,Universit\"at Regensburg, D-93040 Regensburg, Germany}

\author{Endre T\'ov\'ari}
\affiliation{Department of Physics, Budapest University of Technology and Economics and Condensed Matter Research Group of the Hungarian Academy of Sciences, Budafoki ut 8, 1111 Budapest, Hungary}

\author{Markus Weiss}
\affiliation{Department of Physics, University of Basel, Klingelbergstrasse 82, CH-4056 Basel, Switzerland}

\author{Romain Maurand}
\affiliation{University Grenoble Alpes, CEA-INAC-SPSMS F-38000, Grenoble, France}

\author{Klaus Richter}
\affiliation{Institut f\"ur Theoretische Physik,Universit\"at Regensburg, D-93040 Regensburg, Germany}

\author{Christian Sch\"onenberger}
\affiliation{Department of Physics, University of Basel, Klingelbergstrasse 82, CH-4056 Basel, Switzerland}

\date{\today / accepted for publication in Nat. Commun.}

\maketitle


\textbf{Snake states are trajectories of charge carriers curving back and forth along an interface. There are two types of snake states, formed by either inverting the magnetic field direction or the charge carrier type at an interface. Whereas the former has been demonstrated in GaAs-AlGaAs heterostructures, the latter has become conceivable only with the advance of ballistic graphene where a gapless p-n interface governed by Klein tunneling can be formed. Such snake states were hidden in previous experiments due to limited sample quality. Here we report on magneto-conductance oscillations due to snake states in a ballistic suspended graphene p-n-junction which occur already at a very small magnetic field of $20\;$mT. The visibility of $30\%$ is enabled by Klein collimation. Our finding is firmly supported by quantum transport simulations. We demonstrate the high tunability of the device and operate it in different magnetic field regimes.}

\begin{figure}[htbp]
    \centering
      \includegraphics[width=1\textwidth]{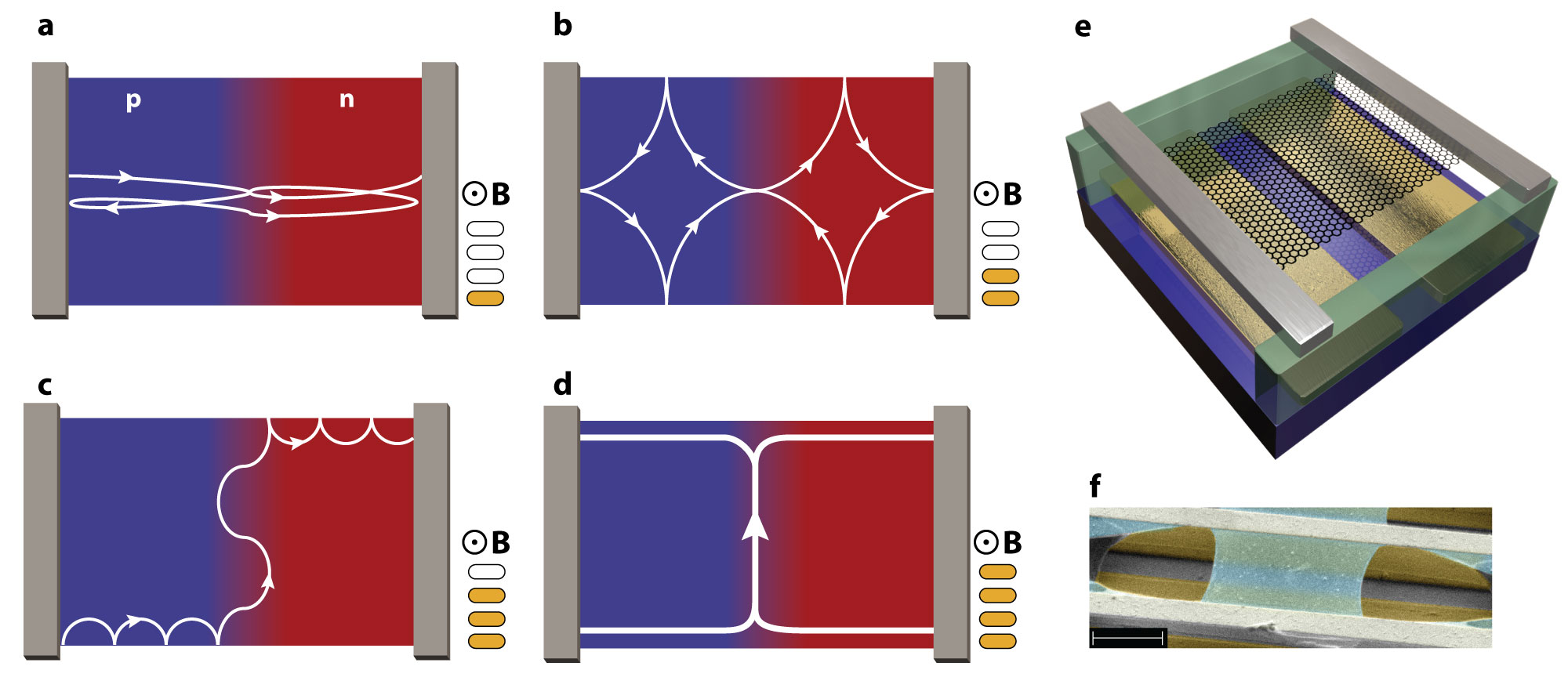}
    \caption{
	\textbf{Evolution of electron states in increasing magnetic field and design of a graphene p-n junction.}
      \textbf{a}, A two-terminal graphene device consisting of a hole (blue) and an electron (red) cavity is sketched. By applying a weak field, the electron trajectories in the p- and n-cavities bend, leading to dispersing Fabry-P\'erot resonances.
	\textbf{b}, The field is increased until the cyclotron orbit becomes comparable to the cavity size, where resonant scar states can occur.
	\textbf{c}, The field is further increased and transport is still described by quasiclassical cyclotron orbits. Snake states along the p-n interface form.
	\textbf{d}, Finally, quantum Hall edge states propagate in opposite direction in the p- and n-region at higher fields.
      	\textbf{e}, 3D design of the measured device. The SiO$_2$ substrate is colored in blue and the bottom-gates in gold. The contacts, supported by the lift-off resist LOR (green) are colored in gray.
      	\textbf{f}, Scanning electron microscope image of a device similar to the measured sample. The graphene is colored in blue and the bottom-gates in gold. The scale bar corresponds to $1\:\mu$m.
}
    \label{fig:figure1}
\end{figure}

A magnetic field  fundamentally modifies the transport properties of an electronic conductor by acting on its charge carriers via the Lorenz force. The most prominent magnetotransport effect is the quantum Hall effect in a two-dimensional electron gas (2DEG). A strong perpendicular magnetic field forces the charge carriers into one-dimensional conduction channels that flow along the edges of a sample. 
At moderate magnetic fields however electron trajectories can be understood in a quasiclassical picture where the Lorenz force bends charge carriers into cyclotron orbits. At the boundary of a conductor charge carriers are not localized and can propagate via so-called skipping orbits.
Magnetic focusing experiments\cite{vanHouten1989, Taychatanapat2013} represent a direct proof of the skipping orbit picture. In such experiments an increase of conductance is observed if the distance between two contacts is an integer multiple of the diameter of a cyclotron orbit. One condition for the observation of such trajectories is ballistic transport over the relevant device dimensions. This has limited the observation of skipping orbits to the cleanest available semiconductor samples.

Since 2004 graphene as a new two-dimensional conductor has moved into the focus of condensed matter research and its behaviour in magnetic field has been intensely investigated. The quality of graphene devices has improved over the recent years and ballistic transport over distances of several microns have been demonstrated recently \cite{Rickhaus2013, Grushina2013}. Since graphene is a gap-less semiconductor, it offers the possibility of creating internal interfaces with opposite charge carrier polarity. These so-called p-n interfaces\cite{Huard2007,Ozyilmaz2007,Gorbachev2008}  are formed by local electrostatic gating.

If electrons that propagate via skipping orbits encounter such a p-n interface, they will turn into snake states. These states consist of alternating half circles with opposite chirality and they transport current along the interface. Similar snake states have first been realized in GaAs/AlGaAs 2DEGs by defining regions of alternating magnetic field direction \cite{Ye1995}. These states share the condition of commensurability (similar to the above described magnetic focusing experiments) with p-n snake states but they do not propagate along a single and tunable interface. Snake states in graphene p-n junctions were claimed to have been observed in disordered substrate supported samples\cite{Williams2011} but the experiment lacked of direct evidence for snaking trajectories.

In this article we investigate ballistic transport across a graphene p-n junction in different magnetic field regimes and identify magneto-conductance oscillations as a direct signature of snake states. These findings are supported by detailed tight-binding simulations that allow us to visualize the alternating cyclotron orbits of the snake states.

\section{results}
\subsection{Evolution of electron trajectories in graphene p-n junction}
In figure \ref{fig:figure1}a-d we illustrate schematically how trajectories under increasing perpendicular magnetic field evolve in such a device. Figure \ref{fig:figure1}a describes the low-field situation where transport is still dominated by Fabry-P\'erot oscillations with slightly bent trajectories \cite{Young2009}. As the field is increased resonant scar states (fig. \ref{fig:figure1}b) may occur, as observed in semiconductor quantum dots \cite{Bird1999}. 
At higher fields (fig. \ref{fig:figure1}c) snake states at the p-n interface govern the electronic properties. 
Finally (fig. \ref{fig:figure1}d) the system enters the quantum Hall regime \cite{Novoselov2005,Zhang2005} where transport is dominated by edge states and Landau level mixing at the p-n interface can occur \cite{Williams2007,Abanin2007}. Even though this article is focusing on snake states, we will discuss the phenomenology of the mentioned magnetic field regimes. By doing so we present an integral picture of graphene p-n physics in magnetic field.

\subsection{Device architecture}

Figure \ref{fig:figure1}e shows the design of the measured device and figure \ref{fig:figure1}f a scanning electron microscope picture of a similar sample. In a suspended $2\times 2\: \mu$m graphene sheet a p-n junction is formed by applying different voltages on the left ($V_{\rm left}$) and right ($V_{\rm right}$)  bottom-gates, resulting in different charge carrier concentrations $n_{\rm left}$ and $n_{\rm right}$. The fabrication follows partly ref. \cite{Tombros2011} which we combined with a wet transfer process allowing us to align the graphene with the bottom-gates. For details see ref.~\cite{Maurand2014486} and Methods.

\begin{figure}[htbp]
    \centering
      \includegraphics[width=1\textwidth]{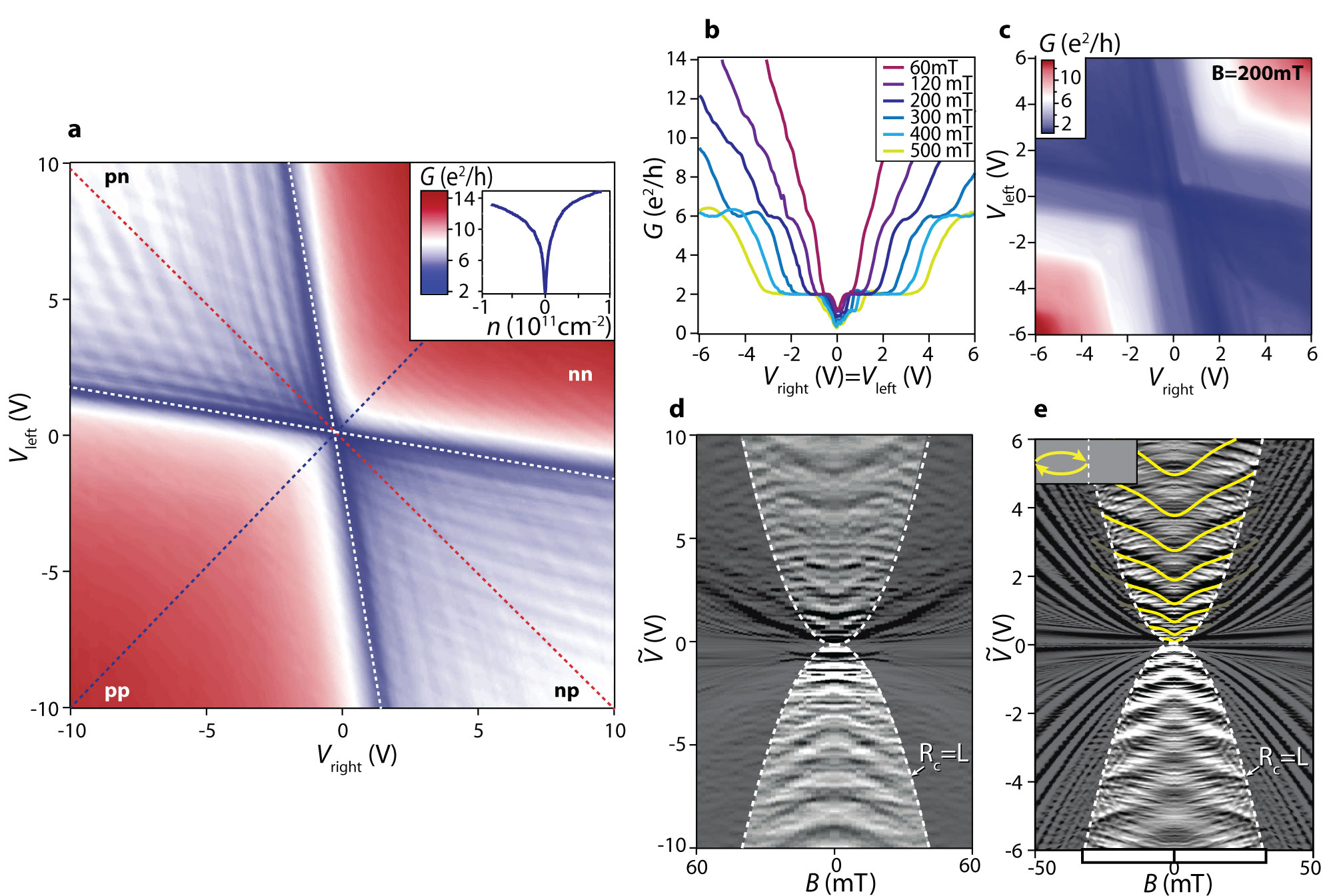}
    \caption{
	\textbf{Device characterization in the Fabry-P\'erot and in the quantum Hall regime.}
     	 \textbf{a}, Two-terminal conductance as a function of left and right gate voltage shows regular Fabry-P\'erot oscillations at zero magnetic field. The inset reveals the narrow Dirac dip along the pp-nn diagonal (blue dashed line) from which a mobility of  $\mu\approx 470\cdot10^3$ cm$^2$V$^{-1}$s$^{-1}$ is deduced.
	 \textbf{b}, Cuts along the same diagonal at different magnetic field strengths exhibit the expected quantum Hall plateaus at $2, 6, 10\:e^{2}/h$. The $G=2\:e^2/h$ plateau is already visible  at $60$ mT.
 	\textbf{c}, The color plot as a function of $V_{\rm left}$ and $V_{\rm right}$ at $200$ mT shows quantum Hall plateaus in the unipolar region.
 	\textbf{d}, The numerical derivative $dG/d\tilde V$ (in arbitrary unit) is recorded as a function of gate voltage $\tilde V$ and $B$ displaying the dispersion of the Fabry-P\'erot oscillations. $\tilde V$ is the magnitude of gate voltage in the situation of antisymmetric charge densities  $n_{\rm left}=-n_{\rm right}$ (red dashed line in \textbf{a}. i.e. the np-pn diagonal). The white dashed curve indicates the line along which the cyclotron radius $R_c$ is equal to the cavity length $L=0.8\:\mu$m, the region $R_c<L$ is darkened and will be discussed in the main text.
	 \textbf{e}, The measured pattern is reproduced by a tight-binding quantum transport calculation based on the designed geometry of the measured device. The small inset shows a resonant electron trajectory at low magnetic field. Constructive interference occurs if the phase along this trajectory is an integer of $2\pi$, leading to the numerical solution of the yellow lines (see Methods).
		}
    \label{fig:figure2}
\end{figure}

\subsection{Measurements in the Fabry-P\'erot and quantum Hall regime}

In the following we characterize the measured device in the zero, low- and high-field regimes. Figure \ref{fig:figure2}a shows a two-dimensional color map of the electrical conductance $G(V_{\rm right},V_{\rm left})$. As soon as a p-n interface is formed, $G$ is lowered drastically. Regular Fabry-P\'erot resonances in the left/right cavity are visible as oscillations perpendicular to the zero density line in the left/right cavity (horizontal/vertical white dashed line in fig. \ref{fig:figure2}a), indicating ballistic transport \cite{Rickhaus2013, Grushina2013}. The inset in figure \ref{fig:figure2}a is a slice along the pp-nn diagonal (blue dashed). We estimate the mobility to be $\mu=\frac{1}{e}\frac{d\sigma}{dn}\approx 470\cdot10^3$ cm$^2$V$^{-1}$s$^{-1}$ at carrier density $n=1.1\cdot10^9$ cm$^{-2}$, which we calculated by a simple parallel plate capacitor model. The mobility is mainly limited by scattering at the contacts \cite{Cayssol2009, Rickhaus2013}.

In figure \ref{fig:figure2}b we show cuts along the pp-nn diagonal at different magnetic fields $B$ and obtain quantum Hall plateaus at $G=G_0\cdot \nu$, where $\nu=2,6,10,... $ is the filling factor \cite{Novoselov2005, Zhang2005} and $G_0=e^2/h$ is the conductance quantum. Even at fields as low as $60$ mT, the $\nu=2$ plateau is visible. The colorscale map in figure \ref{fig:figure2}c taken at $200$ mT shows that plateaus develop in the unipolar region with conductance values given by the lowest number of edge modes in the left or right cavity, i.e. $G=G_{0}\cdot \min(\nu_{\rm right},\nu_{\rm left})=2,6,10\:e^{2}/h$. This observation compares well to experiments of ref. \cite{Williams2007}, even though we apply only $0.2$ T instead of $4$ T. In the bipolar region the conductance stays well below $2\:e^{2}/h$ due to the smoothness of the p-n interface.

In a further step we study the dispersion of the Fabry-P\'erot interference pattern in low magnetic field \cite{Young2009}. Figure \ref{fig:figure2}d shows the numerical derivative $dG/d\tilde V$ as a function of $B$ and $\tilde V$, where $\tilde V$ represents the magnitude of gate voltage in the situation of antisymmetric charge density (red dashed line in fig. \ref{fig:figure2}a i.e. the np-pn diagonal). In this configuration the device consists of two Fabry-P\'erot cavities of equal length $L\approx 0.8\;{\mu\rm m}$. The darkened region where the cyclotron radius $R_c<L$ will be discussed later. In figure \ref{fig:figure2}e a tight-binding transport calculation is shown which reproduces the measured interference pattern very well (for details see ref. \cite{Liu2014} and Methods). We highlight the quality of the measured graphene and the ability of the simulation to capture the complex oscillation pattern of this micron-sized system in magnetic field. The dispersion of the Fabry-P\'erot oscillations can be described by bent electron trajectories such as the one sketched in the inset of figure \ref{fig:figure2}e \cite{Young2009, Rickhaus2013}. The condition for constructive interference is met if the accumulated phase along such a trajectory is a multiple of $2\pi$ (see ``Methods''). The yellow lines in figure~\ref{fig:figure2}e are numerical solutions based on such a condition.

\begin{figure}[htbp]
    \centering
      \includegraphics[width=1\textwidth]{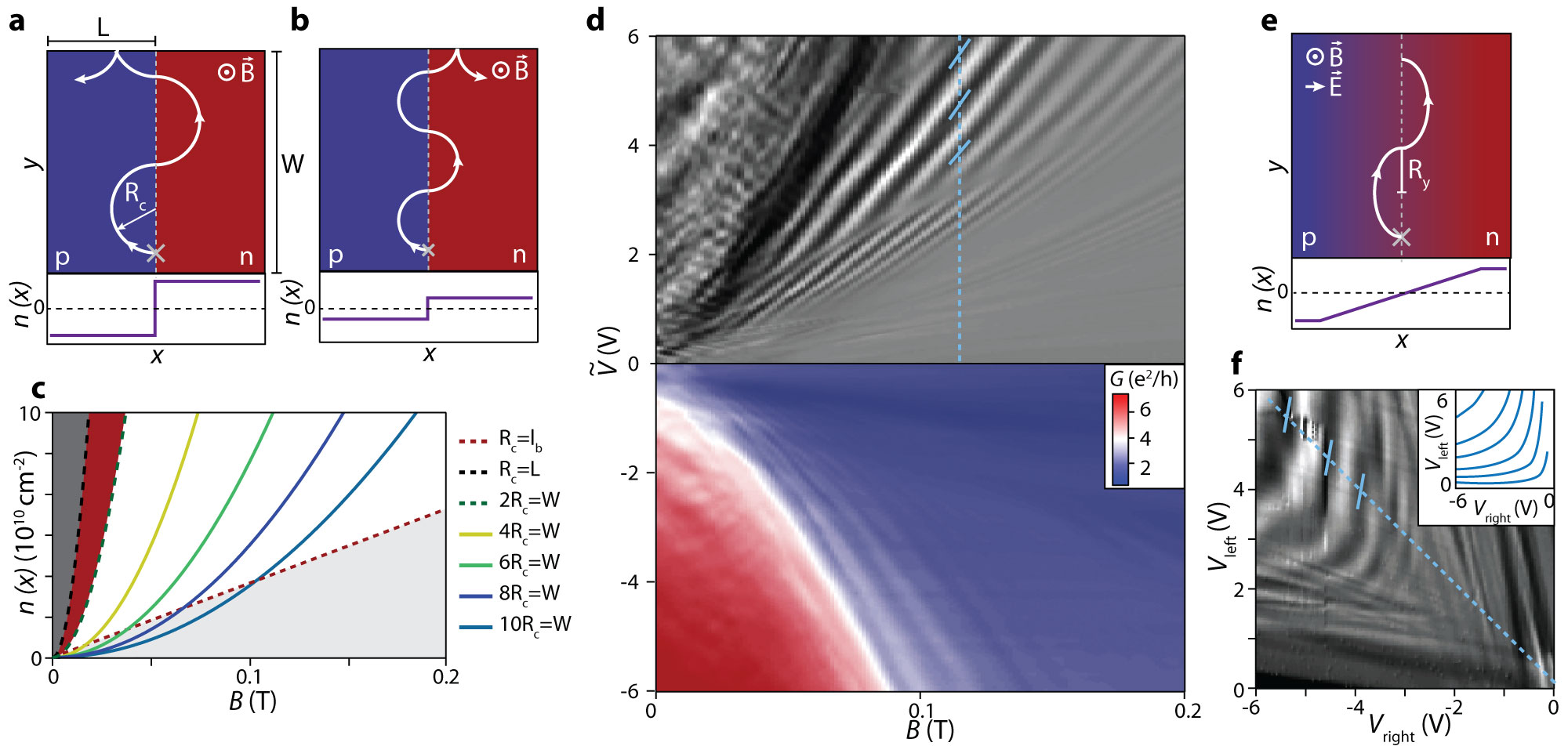}
    \caption{
	\textbf{Parabolic-like conductance oscillations as a signature of snake states.}
	\textbf{a}, Charge carrier trajectory (white) along a sharp p-n junction in perpendicular magnetic field starting at the gray cross, where $R_c$ is the cyclotron radius and $n(x)$ is the electron density.
	\textbf{b}, At lower p- and n-density $R_c$ is reduced. In contrast to \textbf{a}, the trajectory results in current flow towards the right contact.
	\textbf{c}, Curves of constant $R_c$ as a function of $n$ and $B$. The continuous lines are given by the condition that the cyclotron diameter $2R_c$ is commensurate with the sample width $W=2\;\mu$m. Snake states occur between the green dashed $2R_c=W$ and the red dashed $R_c=l_B$ line. The black dashed line indicates up to which field transport is dominated by bent Fabry-P\'erot patterns (dark gray area). In the red area scar states can occur. Below the red dashed line $R_c$ is smaller than the magnetic length $l_B$ and Landau-levels start to dominate the transport (light gray area).
	\textbf{d}, Conductance as a function of antisymmetric gate tuning $\tilde V$ and magnetic field is shown in the lower panel, and its derivative $dG/dB$ (in arbitrary unit) in the upper panel. Striking lines of high and low conductance with a parabolic like B-dependence can be observed. 
	      	\textbf{e}, In a smooth p-n junction (here: linear $n(x)$) the cyclotron orbits become elongated along the $y$-direction due to the additional electric field caused by the density gradient.
      \textbf{f}, $dG(V_{\rm right},V_{\rm left})/d\tilde V$ at $120$ mT. The blue dashed lines in \textbf{d} and \textbf{f} are equivalent. The inset shows lines of constant electric field $\mathbf{E}_x$ at the p-n interface as a function of $V_{\rm right},V_{\rm left}$ taken from electrostatic simulations.
}
   \label{fig:figure3}
\end{figure}

\subsection{Measurement of snake states}

We now discuss the regime where snake states emerge. In figure \ref{fig:figure3}a a snake state at a sharp p-n junction is sketched. Consider a charge carrier trajectory starting at the gray cross with momentum $\mathbf{k}$ in $-x$ direction. Due to the magnetic field the trajectory is bent towards the p-n interface within the cyclotron diameter $2R_c=2\hbar k/eB=2\hbar \sqrt{n\pi}/eB$. If the trajectory hits the p-n interface, the hole will be transmitted to the n side with high probability due to Klein tunneling \cite{Cheianov2006}. At the upper edge of the sample, the snake trajectory scatters at the left side, resulting in a current towards the left contact. At lower $n$, $R_c$ is reduced. As sketched in figure \ref{fig:figure3}b, the snake trajectory scatters to the right at the upper edge, resulting in a net current towards the right contact. With this mechanism one expects conductance oscillations that depend on $B$ and $n$, and constant conductance along curves where $R_c$ is constant. In figure \ref{fig:figure3}c we display calculated functions $n(B)$ for constant $R_c$. Snake states occur once the cyclotron diameter $2R_c$ is smaller than the sample width $W$ (green dashed curve) and can be described by quasiclassical trajectories as long as $R_c$ is larger than the magnetic length $l_B=\sqrt{\hbar /eB}$ (red dashed line). In the regime $W>2R_c>l_B$ additional parabolic lines show the condition for which the number of oscillations in the snake pattern is fixed and commensurate with $W$, i.e. $m\cdot 2R_c = W$ with $m=1,2,3,...$.

In figure \ref{fig:figure3}d we show the measured conductance $G(B,\tilde V)$ (bottom) and its numerical derivative $dG/dB$ (top). The measurement exhibits strong oscillations that follow parabola-like curves. We notice that the oscillations occur on a background of strongly decreasing conductance from $G\approx6\:e^2/h$ to $G<2\:e^2/h$. The steep decrease indicates that the transport becomes dominated by the low density region close to the p-n interface and this happens when $2R_c<W$.

In a real p-n interface the density does not sharply jump from the p to the n side but evolves smoothly. An electron trajectory in such a smooth p-n interface is sketched in figure \ref{fig:figure3}e. Here the density gradient leads to an electric field $\mathbf{E}_x$ and its interplay with the perpendicular magnetic field results in the so-called ExB drift (here along y),  leading to elongated cyclotron orbits. The condition ${R_{\rm y}=const.}$ can be studied in the measurement of figure \ref{fig:figure3}f where we show $dG(V_{\rm right},V_{\rm left})/d\tilde V$ at $B=120\;$mT. The measured interference pattern follows curves of constant $\mathbf{E}_x$ at the p-n interface (obtained from electrostatic simulations) as shown in the inset.

\begin{figure}[htbp]
    \centering
      \includegraphics[width=1\textwidth]{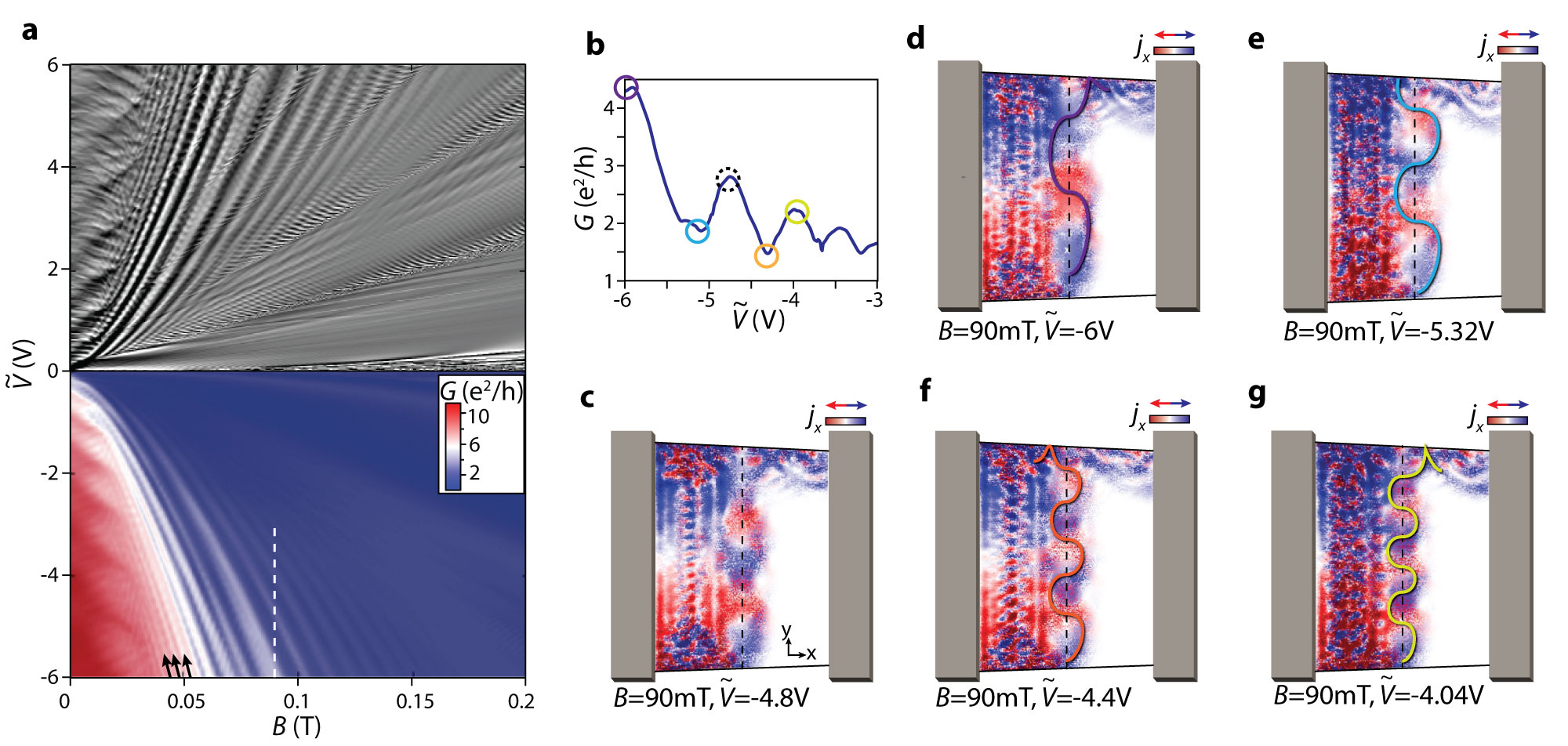}
    \caption{
	\textbf{Tight-binding transport calculation reproducing the experimental results and local current density profiles revealing the snake states.}
      \textbf{a}, Tight-binding calculation of conductance $G(B, \tilde V)$ (bottom) and its numerical derivative $dG/dB$ (top, in arbitrary unit). The parabola-like lines seen in experiment are well captured.
	\textbf{b}, $G(\tilde V)$ along the white dashed line in \textbf{a} at $B=90$ mT.
      \textbf{c}, Calculated $x$-component of the local current density distribution, $j_x$, for electrons injected from the left contact with a small DC-offset at $B=90$ mT and $\tilde V=-4.8$ V (dashed circle in \textbf{b}). The complex resonance pattern in the left cavity consists of ``bubbling'' trajectories that do not contribute to conductance. At the p-n interface (dashed line) an alternating current pointing to left (red) and right (blue) is observed.
      \textbf{d-g}, Current density profiles at different $\tilde V$ corresponding to the circles in \textbf{b}. As a guide to the eye, a snaking trajectory following $j_x$ is added. From one conductance maximum to the next (i.e. \textbf{c} to \textbf{g}) one snake period is added. The snake state in \textbf{e} and \textbf{f} corresponds to a conductance minimum and \textbf{c,d} and \textbf{g} to a maximum.
		}
    \label{fig:figure4}
\end{figure}

\subsection{Tight-binding theory of snake states}

So far we have seen that the oscillation pattern occurs in the regime where snake states are expected (i.e. $2R_c<W$) and that the oscillations are related to transport along the p-n interface. We now present a quantitative comparison between experiment and theory. Figure \ref{fig:figure4}a shows a quantum transport simulation of $G(B,\tilde V)$ and $dG(B,\tilde V)/d\tilde V$ based on a scalable tight-binding model \cite{Liu2014} that fully takes into account the device geometry. The simulation compares very well to the measurement shown in figure \ref{fig:figure3}d. The parabola-like patterns are reproduced and a similarly steep decrease of conductance is obtained.
In figure \ref{fig:figure4}b a slice following the white dashed line in figure \ref{fig:figure4}a is shown. The visibility $\Delta G/G$ of the oscillations reaches $30\%$ in theory and experiment and is enabled by the strong Klein collimation at the smooth p-n-interface.

Next we apply the Keldysh-Green's function method to extract local current density profiles (see Methods) at high and low conductance along this line. In figure \ref{fig:figure4}c we show the $x$-component of the current density $j_x$, taken at $\tilde V=-4.8$ V (dashed circle in fig. \ref{fig:figure4}b). The current is injected from the left contact using a small DC offset. In the left cavity a complex resonance pattern appears, given by so-called "bubbling" trajectories \cite{Carmier2011} which are reflected before reaching the p-n interface and do not contribute to current between the contacts. The pattern relevant for transport is located at the p-n interface (dashed line). We observe that $j_x$ changes sign along the p-n-interface and that the blue and red regions penetrate the dashed line, indicating that transport is dominated by Klein-collimated snake trajectories.
As a guide to the eye we added a curve in figures \ref{fig:figure4}d-g that follows the snake state. This is done for different $\tilde V$ values for which $G$ is maximal/minimal at a fixed magnetic field of $90\;$mT (colored circles in fig. \ref{fig:figure4}b). In figure \ref{fig:figure4}d for example, the current density profile corresponds to a conductance maximum where the current points to the right at the upper edge of the p-n interface. One  period is added by changing $\tilde V$ from one maximum to the next. More current density profiles evolving with $\tilde V$ at fixed $B=90$ mT are shown in Supplementary Movie 1. By tracing along one of the parabolic-like patterns, the current density profile of the snake state stays constant (an example is given in Supplementary Movie 2).

The conductance oscillates as a function of the ratio $W/2R_y$, the exact snake period corresponding to $4R_y$ is, however, difficult to determine using quasiclassical trajectories, since current is injected from many points under various angles resulting in a complex cusp structure similar to what was predicted in refs \cite{Davies2012, Patel2012, Milovanovic2013, Milovanovic2014}. The excellent agreement between measurement and calculated conductance for which we could determine local current density profiles clearly indicates that the oscillations  result from snake state trajectories.

There are additional parabola-like structures at lower magnetic field indicated by arrows in figure \ref{fig:figure4}a, these structures are however less pronounced in the experimental data. Those resonances occur in a regime where scar states (fig. \ref{fig:figure1}b) would be expected. The parabola-like behaviour indicates that the states are commensurate with a cavity dimension. In the model, the resonances disappear for non-reflective contacts as expected for scar states.

\section{Discussion}

We investigated magneto-conductance of a ballistic graphene p-n-junction in the different magnetic field regimes. We have observed resonance patterns occurring in the intermediate quasi-classical regime in experiment and theory which result from the formation of snake states at the p-n interface.  Among many other possibilities these states can be used to guide electrons on arbitrary paths with a high efficiency even at very low magnetic fields. This could be used to guide electrons away from sample edges to suppress uncontrolled momentum or spin-scattering. The directional scattering at the sample boundaries could be used to implement multi-terminal switches \cite{Chen2012a,Milovanovic2013}. Furthermore, the similarity between Andreev reflection and Klein tunneling is stressed in theory \cite{Beenakker2008} leading to a correspondence of snake states and Andreev edge states which are of theoretical \cite{Hoppe2000} and experimental \cite{Rickhaus2012} interest. Our work points out that snake states are highly tunable, occur at low fields and that ballistic graphene p-n-junctions in magnetic field reveal novel and intriguing phenomena.

\newpage


\section{Methods}

\emph{Experimental methods.} High quality graphene is obtained by in situ current annealing \cite{Moser2007}. All measurements were performed in a variable temperature Helium cryostat with a base temperature of $1.5$ K. We measured differential conductance $G=dI/dV$ by standard lock-in technique applying an AC voltage of $0.1$ mV at $77$ Hz.

For the quantum Hall data of figure \ref{fig:figure2}b-c we subtracted a contact resistance of $1.2$ k$\Omega$.

We extracted the cavity length $L$ used in figure \ref{fig:figure2}d-e from the spacing $\Delta n$ between resonant Fabry-P\'{e}rot peaks in the bipolar situation $\Delta n=2\sqrt{\pi n}/L$ and obtained $L\approx0.8\:\mu$m.

\emph{Simulation methods.}
Real-space Green's function method in the tight-binding framework using a scaled graphene Hamiltonian \cite{Liu2014} is applied to simulate ballistic quantum transport in the present device, taking into account the realistic on-site energy profile obtained by three-dimensional electrostatic simulation for the self-partial capacitances of the bottom gates. All the presented conductance simulations are obtained by calculating the transmission function at zero temperature, combined with the contact resistance $1.2~\mathrm{k\Omega}$. Local current densities are imaged by applying the Keldysh-Green's function method in the linear response regime \cite{Cresti2008a} based on the same model Hamiltonian used for conductance simulation. At each lattice site $n$, the bond charge current density $\mathbf{J}_n=\sum_m\mathbf{e}_{n\rightarrow m}\langle J_{n\rightarrow m}\rangle$ is computed, where the summation runs over all the sites $m$ nearest to $n$, $\mathbf{e}_{n\rightarrow m}$ is the unit vector pointing from $n$ to $m$, and $\langle J_{n\rightarrow m}\rangle$ is the quantum statistical average of the bond charge current operator $J_{n\rightarrow m}$ \cite{Nikolic2006}. After computing for each site, the position-dependent current density profile $\mathbf{J}(x,y)=[j_x(x,y),j_y(x,y)]$ is imaged. In fig. \ref{fig:figure4}c-g the $x$-component $j_x(x,y)$ is shown.

The low-field Fabry-P\'erot interference contours sketched in Fig.\ 2e are numerically obtained from solving the resonance condition $\Delta\Phi=2j\pi$, arising from the path difference between the directly transmitted and twice reflected trajectories within the p cavity as sketched in the inset of Fig.\ 2e, which is found to be the major interference contribution. For such a simplified model the phase difference is given by $\Delta\Phi=\Phi_{\rm WKB}+\Phi_{\rm AB}+\Phi_{\rm Berry}+\Phi_0$, where $\Phi_{\rm WKB}=\sqrt{k_F^2-(eBL/2\hbar)^2}\cdot 2L$ is the kinetic WKB-phase, $\Phi_{\rm AB}=\delta A\cdot eB/\hbar$ is the Aharanov-Bohm phase due to the flux enclosed by the bent orbit segments, $\Phi_{\rm Berry}=-\pi(1-e^{-(B/B_c)^2})$ is the Berry phase, and $\Phi_0=\pi$ is a constant phase due to reflections off the two p-n junction interfaces of the p cavity (smooth at middle and sharper at the contact side). Here $k_F$ is the numerical average of $\sqrt{\pi|n(x,y=0)|}$ within the p cavity. The cavity size $L$ is numerically determined and is about half of the flake length $L/2=840~{\rm nm}$. The loop area is given by $\delta A=R_c^2(\phi-\sin\phi)$ with $\phi=2\arcsin(L/2R_c)$ and $R_c=\hbar k_F/eB$. The form of the Berry phase follows from the consideration of \cite{Shytov2008} with the critical field estimated by $B_c=(2\hbar k_F/eL)\sqrt{1-T_c}$, where the critical transmission value $T_c$ is a parameter close to one and does not significantly influence the shape of the contours; $T_c=0.95$ is chosen. The contours sketched in Fig.\ 2e correspond to $j=1,2,\cdots,8$.

\textbf{Acknowledgments}
M.-H.L. and K.R. acknowledge financial support by the Deutsche Forschungsgemeinschaft (SFB 689 and SPP 1666). This work was further funded by the Swiss National Science Foundation, the Swiss Nanoscience Institute, the Swiss NCCR QSIT, the Graphene Flagship, the ESF programme Eurographene, the EU FP7 project SE2ND, the ERC Advanced Investigator Grant QUEST, and the EU flagship project graphene.

The authors thank Andreas Baumgartner, Laszlo Oroszlany and Szabolcs Csonka for fruitful discussions.

\textbf{Author contributions}
P.R., P.M. and E.T. fabricated the devices. Measurements were performed by P.R., P.M., R.M. and M.W. M.-H.L. performed simulations. C.S. and K.R. guided the work. All authors worked on the manuscript.

\newpage

\bibliographystyle{unsrt}
\bibliography{pnB_paper}

\begin{thebibliography}{10}

\bibitem{vanHouten1989}
{van Houten H}, Beenakker, Williamson, Broekaart, {van Loosdrecht PH}, {van
  Wees BJ}, Mooij, Foxon, and Harris.
\newblock Coherent electron focusing with quantum point contacts in a
  two-dimensional electron gas.
\newblock {\em Phys Rev B Condens Matter}, 39(12):8556--8575, 1989.

\bibitem{Taychatanapat2013}
Thiti Taychatanapat, Kenji Watanabe, Takashi Taniguchi, and Pablo
  Jarillo-Herrero.
\newblock Electrically tunable transverse magnetic focusing in graphene.
\newblock {\em Nat Phys}, 9(4):225--229, April 2013.

\bibitem{Rickhaus2013}
P.~Rickhaus, R.~Maurand, M.-H. Liu, M.~Weiss, K.~Richter, and
  C.~Sch\"{o}nenberger.
\newblock Ballistic interferences in suspended graphene.
\newblock {\em Nature Comm}, 4:2342, 2013.

\bibitem{Grushina2013}
A.~L. Grushina, D.-K. Ki, and A.~F. Morpurgo.
\newblock A ballistic pn junction in suspended graphene with split bottom
  gates.
\newblock {\em Appl. Phys. Lett.}, 102:223102, 2013.

\bibitem{Huard2007}
B.~Huard, J.~A. Sulpizio, N.~Stander, K.~Todd, B.~Yang, and
  D.~Goldhaber-Gordon.
\newblock Transport measurements across a tunable potential barrier in
  graphene.
\newblock {\em Phys. Rev. Lett.}, 98:236803, 2007.

\bibitem{Ozyilmaz2007}
B.~Ozyilmaz, P.~Jarillo-Herrero, D.~Efetov, D.~A. Abanin, L.~S. Levitov, and
  P.~Kim.
\newblock Electronic transport and quantum hall effect in bipolar graphene
  p-n-p junctions.
\newblock {\em Phys. Rev. Lett.}, 99:166804, 2007.

\bibitem{Gorbachev2008}
R.~V. Gorbachev, A.~S. Mayorov, A.~K. Savchenko, D.~W. Horsell, and F.~Guinea.
\newblock Conductance of p-n-p graphene structures with "air-bridge" top gates.
\newblock {\em Nano Lett}, 8:1995--1999, 2008.

\bibitem{Ye1995}
P.D. Ye, D.~Weiss, and R.R. Gerhardts.
\newblock Electrons in a periodic magnetic field induced by a regular array of
  micromagnets.
\newblock {\em Phys. Rev. Lett.}, 74:3013, 1995.

\bibitem{Williams2011}
J.~R. Williams and C.~M. Marcus.
\newblock Snake states along graphene p-n junctions.
\newblock {\em Phys. Rev. Lett.}, 107:046602, 2011.

\bibitem{Young2009}
Andrea~F. Young and Philip Kim.
\newblock Quantum interference and klein tunnelling in graphene
  heterojunctions.
\newblock {\em Nat. Phys.}, 5:222--226, 2009.

\bibitem{Bird1999}
J.~Bird, R.~Akis, D.~Ferry, D.~Vasileska, J.~Cooper, Y.~Aoyagi, and T.~Sugano.
\newblock {Lead-Orientation-Dependent Wave Function Scarring in Open Quantum
  Dots}.
\newblock {\em Phys. Rev. Lett.}, 82:4691--4694, 1999.

\bibitem{Novoselov2005}
K.~S. Novoselov, A.~K. Geim, S.~V. Morozov, D.~Jiang, M.~I. Katsnelson, I.~V.
  Grigorieva, S.~V. Dubonos, and A.~A. Firsov.
\newblock Two-dimensional gas of massless dirac fermions in graphene.
\newblock {\em Nature}, 438:197--200, 2005.

\bibitem{Zhang2005}
Y.~Zhang, Y.-W. Tan, H.~L. Stormer, and P.~Kim.
\newblock Experimental observation of the quantum hall effect and berry's phase
  in graphene.
\newblock {\em Nature}, 438:201, 2005.

\bibitem{Williams2007}
J.~R. Williams, L.~Dicarlo, and C.~M. Marcus.
\newblock Quantum hall effect in a gate-controlled p-n junction of graphene.
\newblock {\em Science}, 317:638--641, 2007.

\bibitem{Abanin2007}
D.~A. Abanin and L.~S. Levitov.
\newblock Quantized transport in graphene p-n junctions in a magnetic field.
\newblock {\em Science}, 317:641, 2007.

\bibitem{Tombros2011}
N.~Tombros, A.~Veligura, J.~Junesch, J.J. van~den Berg, P.J. Zomer, I.J.
  Vera-Marun, H.T. Jonkman, and B.~van Wees.
\newblock Large yield production of high mobility freely suspended graphene
  electronic devices on a polydimethylglutarimide based organic polymer.
\newblock {\em J. Appl. Phys}, 109:093702, 2011.

\bibitem{Maurand2014486}
R.~Maurand, P.~Rickhaus, P.~Makk, S.~Hess, E.~Tovari, C.~Handschin, M.~Weiss,
  and C.~Schoenberger.
\newblock Fabrication of ballistic suspended graphene with local-gating.
\newblock {\em Carbon}, 79:486, 2014.

\bibitem{Cayssol2009}
J.~Cayssol, B.~Huard, and D.~Goldhaber-Gordon.
\newblock Contact resistance and shot noise in graphene transistors.
\newblock {\em Phys. Rev. B.}, 79:075428, 2009.

\bibitem{Liu2014}
M-H. Liu, P.~Rickhaus, E.~T\`ovari, R.~Maurand, F.~Tkatschenko, M.~Weiss,
  C.~Sch\"{o}nenberger, and K.~Richter.
\newblock Scalable tight-binding model for graphene.
\newblock {\em arXiv:1407.5620v1, to be published in Phys. Rev. Lett.}, 2014.

\bibitem{Cheianov2006}
V.~Cheianov and V.~Fal’ko.
\newblock {Selective transmission of Dirac electrons and ballistic
  magnetoresistance of n-p junctions in graphene}.
\newblock {\em Physical Review B}, 74(4):041403, 2006.

\bibitem{Carmier2011}
P.~Carmier, C.~Lewenkopf, and D.~Ullmo.
\newblock Semiclassical magnetotransport in graphene n-p junctions.
\newblock {\em Phys. Rev. B.}, 84:195428, 2011.

\bibitem{Davies2012}
Nathan Davies, Aavishkar~a. Patel, Alberto Cortijo, Vadim Cheianov, Francisco
  Guinea, and Vladimir~I. Fal'ko.
\newblock {Skipping and snake orbits of electrons: Singularities and
  catastrophes}.
\newblock {\em Phys. Rev. B.}, 85:155433, 2012.

\bibitem{Patel2012}
A.~A. Patel, N.~Davies, V.~Cheianov, and V.~I. Fal'ko.
\newblock Classical and quantum magneto-oscillations of current flow near a p-n
  junction in graphene.
\newblock {\em Phys. Rev. B.}, 86:081413, 2012.

\bibitem{Milovanovic2013}
S.~P. Milovanovic, M.~{Ramezani Masir}, and F.~M. Peeters.
\newblock {Spectroscopy of snake states using a graphene Hall bar}.
\newblock {\em Applied Physics Letters}, 103:233502, 2013.

\bibitem{Milovanovic2014}
S.~P. Milovanovi\'{c}, M.~{Ramezani Masir}, and F.~M. Peeters.
\newblock {Magnetic electron focusing and tuning of the electron current with a
  pn-junction}.
\newblock {\em Journal of Applied Physics}, 115(4):043719, January 2014.

\bibitem{Chen2012a}
Jiang-chai Chen, X.~C. Xie, and Qing-feng Sun.
\newblock {Current oscillation of snake states in graphene p-n junction}.
\newblock {\em Physical Review B}, 86(3):035429, July 2012.

\bibitem{Beenakker2008}
C.~Beenakker, A.~Akhmerov, P.~Recher, and J.~Tworzydlo.
\newblock {Correspondence between Andreev reflection and Klein tunneling in
  bipolar graphene}.
\newblock {\em Physical Review B}, 77:075409, 2008.

\bibitem{Hoppe2000}
H.~Hoppe, U.~Zulicke, and G.~Schon.
\newblock Andreev reflection in strong magnetic fields.
\newblock {\em Phys. Rev. Lett.}, 84:1804, 2000.

\bibitem{Rickhaus2012}
Peter Rickhaus, Markus Weiss, Laurent Marot, and Christian Sch\"{o}nenberger.
\newblock Quantum hall effect in graphene with superconducting electrodes.
\newblock {\em Nano Lett.}, 12:1942, 2012.

\bibitem{Moser2007}
J.~Moser, A.~Barreiro, and A.~Bachtold.
\newblock {Current-induced cleaning of graphene}.
\newblock {\em Appl. Phys. Lett.}, 91:163513, 2007.

\bibitem{Cresti2008a}
A.~Cresti, G.~Grosso, and G.~Parravicini.
\newblock {Electronic states and magnetotransport in unipolar and bipolar
  graphene ribbons}.
\newblock {\em Phys. Rev. B.}, 77:115408, 2008.

\bibitem{Nikolic2006}
B.~Nikoli\'{c}, L.~Z\^{a}rbo, and S.~Souma.
\newblock {Imaging mesoscopic spin Hall flow: Spatial distribution of local
  spin currents and spin densities in and out of multiterminal spin-orbit
  coupled semiconductor nanostructures}.
\newblock {\em Phys. Rev. B.}, 73:075303, 2006.

\bibitem{Shytov2008}
A.~Shytov, M.~Rudner, and L.~Levitov.
\newblock {Klein Backscattering and Fabry-P\'{e}rot Interference in Graphene
  Heterojunctions}.
\newblock {\em Phys. Rev. Lett.}, 101:156804, 2008.

\end{thebibliography}

\end{document}